\documentstyle[12pt]{article}  
\newcommand{\al}{\alpha} 
\newcommand{\olocl}{\int_{0}^{1}dt\int_{0}^{1}d\al} 
\newcommand{\usual}{{4Q^2 \over M_W^2}t^2 \al (1-\al)+i\epsilon} 
  
\newcommand{\usualp}{{4Q^2 \over M_W^2}t^3(3t-2) \al (1-\al)} 
\begin{document} 
\begin{titlepage} 
\begin{centering} 
{\Large{\bf Trilinear Gauge Boson Couplings in the MSSM}}\\ 
\vspace{1.5cm} 
{{\bf E. N. Argyres }\hspace{.02cm}}$^{a\star}$, 
{{\bf A. B. Lahanas }\hspace{.02cm}}$^{b\dagger}$,  
 {{\bf C. G. Papadopoulos}\hspace{.02cm}}$^{c,d \star\star}$ 
\\                  and  
{\hspace{.2cm}}{{\bf V. C. Spanos }\hspace{.02cm}}$^{c\ddagger}$  \\ 
\vspace{.4cm} 
$^{a}$ NRCPS, `Democritos', Aghia Paraskevi, GR -- 15310 Athens, Greece \\ 
$^{b}$ University of Athens, Physics Department,  
%\vspace{.05in} 
Nuclear and Particle Physics Section,\\ 
%\vspace{.05in} 
GR -- 15771  Athens, Greece\\ 
$^{c}$ University of Durham, Physics Department, Science Laboratories, \\ 
South Road, Durham DH1 3LE, England  \\ 
$^{d}$ CERN, Theory Division, CH-1211, Geneva 23, Switzerland \\ 
\vspace{1.2cm} 
{\bf Abstract}\\ 
\vspace{.1in} 
\end{centering} 
{  
\noindent 
%{\small { 
{{We study the $C$ and $P$ even $WW\gamma$ and $WWZ$ trilinear gauge boson  
vertices (TGV's),  
in the context of the MSSM  assuming that the  
external $W$'s are on their mass shell. We find that for energies 
${\sqrt{s}} \equiv {\sqrt {q^2}} \le 200 \; GeV$  squark and slepton  
contributions to 
the aforementioned couplings are two orders of magnitude smaller than those  
of the Standard Model (SM). In the same energy range the bulk of the  
supersymmetric Higgs corrections to the TGV's is  due to the lightest neutral  
Higgs, $h_0$, whose contribution is like that of a Standard Model Higgs   
of the same mass. 
The 
contributions of the Neutralinos and Charginos are sensitive to the input 
value for the soft gaugino mass $M_{1/2}$, being more pronounced for values 
$M_{1/2} < 100 \; GeV$. In this case and in the unphysical region, 
$0 < \sqrt{s} < 2 M_W $, their contributions are substantially enhanced  
resulting in large corrections to the static quantities of the $W$ boson.  
However, such an enhancement is not observed in the physical region.  
In general for  
$2 M_W < \sqrt{s} < 200 \; GeV $ the MSSM predictions differ from those of  
the SM but they are of the same order of magnitude. 
To be detectable deviations from the SM require sensitivities    
reaching the per mille  
level and hence unlikely to be observed at LEP200. For higher energies  SM and 
MSSM predictions exhibit a fast fall off behaviour, in accord with unitarity 
requirements, getting smaller,in most cases, 
by almost an order of magnitude already at  
energies $\sqrt{s} \approx0.5 \; TeV $. }}}    
\paragraph{} 
\par 
\vspace{.3cm} 
\begin{flushleft} 
%Athens University 
UA/NPPS-18B  \\ 
October 1995\\ 
\end{flushleft}   
%\vspace{0.5cm}     
\rule[0.in]{4.5in}{.01in} \\      
\vspace{.3cm}  
E-mail: 
\parbox[t]{13cm}{ {$^{\star}$argyres@cyclades.nrcps.ariadne-t.gr},
{$^{\dagger}$alahanas@atlas.uoa.ariadne-t.gr}, 
{$^{\star\star}$papadopo@alice.nrcps.ariadne-t.gr},
{$^{\ddagger}$V.C.Spanos@durham.ac.uk}.
 } 
\end{titlepage} 
%%%%%%%%%%%%%%%%%%%%%%%%% 
%\end{flushleft} 
%\rule[0.in]{4.5in}{.01in}\\ 
%E-mail: 
%%%%%%%%%%%%%%%%%%%%%%%%%%%%%%%%%%%%%%%%%%%%%%%% 
%\baselineskip=18pt       
\section*{1. Introduction} 
The Standard Model (SM) has been remarkably successful in describing particle 
interactions at energies around $\sim 100\:GeV$. Precise measurements at   
LEP provided accurate tests of the standard theory of electroweak  
interactions $^{\cite{hollik,schwarz}}$, but we are still lacking a  
direct experimental confirmation  
of the non-abelian structure of the standard theory. The $WW\gamma$, $WWZ$ 
and 
$ZZ\gamma$ couplings are uniquely determined within the context of the SM and 
such couplings will be probed in the near future with high accuracy.  
The study of the trilinear gauge bosons vertices (TGV's) in the   
$e^{+}e^{-}\rightarrow W^{+}W^{-}$ process is the primary motivation for the  
upgrading of LEP200 $^{\cite{treille}}$ and the potential for measuring  
these has been discussed 
in detail $^{\cite{treille,bil,flei}}$.   
So far there are no stringent experimental bounds on these 
couplings $^{\cite{ua2}}$ and the efforts of the various experimental groups  
towards this direction are still continuing.  
Precise measurements of these vertices are 
of vital importance not only for the SM itself but also for 
probing new physics which may be revealed at scales larger than the Fermi 
scale. 
\par 
The gauge boson vertex has been the subject of an intense theoretical study  
during 
the last years. In particular the $WWV$ vertex $(V=\gamma$ or $Z)$ has been 
analysed in detail within the framework of the standard theory, as well as 
extensions of it, and its phenomenology has been discussed. The lagrangian 
density describing the $WWV$ interaction is given by $^{\cite{gounar,falk}}$ 
\begin{eqnarray} 
& {\cal L}^{WWV}=-ig_{WWV}[(W_{\mu\nu}^{\dagger}W^{\mu}V^{\nu}-h.c.)+ 
		 {\kappa}_{V}W_{\mu}^{\dagger}W_{\nu}F^{\mu\nu}+ 
		 \frac{{\lambda}_{V}}{M_{W}^{2}}W_{\lambda\mu}^{\dagger} 
		 W^{\mu}_{\nu} 
		 V^{\nu\lambda}+...\:].
%& g_{WWV}=\left\{ \begin{array}{ll} 
%                   e & \mbox{for $V=\gamma$} \nonumber \\ 
%     e\cot\theta_{W} & \mbox{for $V=Z$,} \nonumber  
%                \end{array} 
%        \right. 
\end{eqnarray} 
In this $g_{WWV}=e$ for $V=\gamma$ and $g_{WWV}=e\cot\theta_{W} $
for $V=Z$.
The ellipsis stand for $P$ or $C$ odd terms and higher  
dimensional operators. 
In Eq. (1) the scalar 
components for all gauge bosons involved have been omitted, that is  
${\partial}{\cdot}W={\partial}{\cdot}V=0$, since essentially they 
couple to massless fermions 
\footnote{For on shell $W$'s we have 
$({\Box}+m_{W}^{2})W_{\mu}=0$ and certainly ${\partial}{\cdot}W=0$, 
while for the photon ${\partial}{\cdot}A$ vanishes on account of current 
conservation.}.  
At the tree level  $\kappa_{V}$ and $\lambda_{V}$  
have the values $\kappa_{V}=1$, $\lambda_{V}=0$. However radiative 
corrections modify these, the order of magnitude of these corrections  
being ${\cal O}(\frac{\alpha}{\pi})\sim10^{-3}$. 
Sensitivity limits of this order of magnitude   
will not be reached at LEP200 but can be 
achieved in future colliders where the TGV's can be studied in detail and  
yield valuable information not only for the self consistency of the SM  
but also for probing underlying new physics. Any new dynamics whose onset 
lies in the $TeV$ range modifies $\kappa_{V}$, $\lambda_{V}$ and 
deviations from the SM predictions are expected.  
 
Supersymmetry (SUSY) is 
an extension of the SM which is theoretically motivated but without any 
experimental confirmation $^{\cite{Nilles}}$.   
The only experimental hint for its existence  
derives from the fact that the gauge couplings unify at energies  
$\sim10^{16}\:GeV$, if we adopt a supersymmetric extension of the SM in which 
SUSY is broken at energies 
$M_{SUSY}{\sim}{\cal O}(1)\:TeV$ $^{\cite{amaldi}}$ . 
Supersymmetric particles with such large masses can be produced in the 
laboratory provided we have very high energies and luminocities.  
However, their existence affects   
$\kappa_{V}$ and $\lambda_{V}$, even at  
energies lower than the SUSY production threshold making them deviate from  
the SM predictions. Therefore the study of these quantities may  
furnish us with a good 
laboratory to look for signals of supersymmetry at energies below the SUSY 
production threshold. In any case such studies serve as a complementary 
test, along with other efforts, towards searching for signals of new physics  
and supersymmetry is among the prominent  candidates.  
\par 
In the SM $\kappa_{V}$ and 
$\lambda_{V}$ as functions of the momentum $q^{2}$ carried by the $V$ 
boson \mbox{$(V=\gamma,Z)$}, for on shell $W$'s, have been studied in  
detail $^{\cite{argy}}$, but  
a similar analysis has not been carried out within the context of the MSSM.  
Only the quantities $\kappa_{\gamma}(q^{2}=0)$, $\lambda_{\gamma}(q^{2}=0)$ 
have been considered $^{\cite{spa}}$,  
which are actually related to the static  
magnetic dipole $(\mu_{W})$ and electric quadrupole $(Q_{W})$ moments 
of the $W$ boson. To be of relevance for future collider experiments 
the form factors $\kappa_{\gamma,Z}$, $\lambda_{\gamma,Z}$ should be 
evaluated in the region $q^{2}>4M_{W}^{2}$. The behavior of 
$\kappa_{\gamma,Z}$, $\lambda_{\gamma,Z}$ in this physical region may 
be different from that at $q^{2}=0$, especially when the energy   
gets closer to $M_{SUSY}$  and supersymmetric particles may yield sizeable 
effects due to the fact that we are approaching their thresholds.  
In those cases an  
enhancement of their corresponding contributions is expected, unlike 
SM  contributions which are suppressed  
in this high energy regime. 
\par 
In this work we undertake this problem and study the $C$ and $P$ even  
$WW\gamma$, $WWZ$ vertices in the context of the MSSM, with radiative 
electroweak symmetry breaking, 
when the external 
$W$ bosons are on their mass shell. Such studies are important in 
view of forthcoming experiments at LEP200 and other future collider  
experiments which will probe the structure of the gauge boson couplings and  
test with high accuracy the predictions of the SM. If deviations from the SM 
predictions are observed these experiments will signal the presence of new  
underlying dynamics at scales larger than the Fermi scale.  
%\par 
% 
%\newpage 
%%%% 
%\vspace{.5cm} 
\vspace{.25mm}
%%%%%%%%%%%%%%%%%%%%%%%%%%%% section 1 %%%%%%%%%%%%%%%%%%%%%%%%%%%%%%% 
\section*{1. The SM contribution to  
$\Delta k_{V}(Q^{2})$, $\Delta Q_{V}(Q^{2})$} 
%%%%% 
Although the SM contributions to the TGV's have already been calculated   
in the literature $^{\cite{argy}}$, for reasons of completeness 
we shall briefly discuss them in this section paying special 
attention to the contributions of fermions and gauge bosons.  
\par 
In momentum space the most general $WWV$ vertex $(V=\gamma$ or $Z)$ with 
the two $W$'s on shell and keeping only the transvers degrees of freedom for  
the $\gamma$ or $Z$ can be writen as $^{\cite{gounar}}$ 
\begin{eqnarray} 
\Gamma_{\mu \alpha \beta}^{V}& = &-ig_{WWV} \, \{ \,   
		     f_{V}[2g_{\alpha \beta}{\Delta_\mu} 
       +4(g_{\alpha \mu}{Q_\beta}-g_{\beta \mu}{Q_\alpha})] \nonumber\\ 
   &   &+2 \, {\Delta k}_V \,  
			  (g_{\alpha \mu}{Q_\beta}-g_{\beta \mu}{Q_\alpha})+ 
       4 \, {\frac  {{\Delta Q}_V}  {M_{ W}^{ 2}}} \, {\Delta_\mu}  
       ({Q_\alpha}{Q_\beta}-{\frac {Q^{ 2}}{2}} g_{\alpha \beta})\}+... 
\end{eqnarray} 
where ${{\Delta k}_V} \, \equiv \, \kappa_V + {\lambda}_V -1 \, , \, 
{{\Delta Q}_V} \, \equiv \, -2 \,  {\lambda}_V $. 
The kinematics of the vertex is shown in  Fig. 1 .  
\par 
The ordinary matter fermion  
contributions at $Q^{2}=0$ have been studied 
elsewhere $^{\cite{bard,couture}}$. However in those works there is an  
important sign error which affects substantially the results given 
in those references $^{\cite{spa}}$. This has also been pointed out  
independently in ref. ${\cite{cula}}$ . 
The  consequences of this  for the static quantities of the $W$ 
boson $\mu_{W}$, $Q_{W}$ has been discussed in detail in ref.  
${\cite{spa}}$. The fermion  
contributions to $\Delta k_{\gamma}(Q^{2}=0)$ are large and 
negative partially cancelling contributions of the gauge bosons and  
standard model Higgs which are positive. 
\par 
At $Q^2 \neq 0$ and in units of $g^2/{16\pi^{2}}$ 
the fermion contributions of the graph shown in Fig. 1,  
when the internal lines are $(P_1,P_2,P_3)=(f,f,f')$, 
are given by, 
%\newpage 
% 
\begin{eqnarray} 
\Delta k_{V}&=&-c_{V} T_3^f  C_g    \olocl  \{g_{L}^{f} 
     [t^4+t^3(-1+R_{f'}-R_f)+t^2(R_f-R_{f'}) \nonumber\\ 
    & &+\frac{4Q^2}{M_W^2}t^3(3t-2)\alpha(1-\alpha)]       
	  +g_R^f [\, R_f \, {t^2}]  \} {1\over L_f^2}  \label{fer1}    \\ 
\Delta Q_{V}&=& -c_{V} T_3^f  C_g \, g_{L}^{f}\int_{0}^{1}dt 
    \int_{0}^{1}d\alpha {8t^3(1-t)(1-\alpha)\alpha \over L_f^2}  \label{fer2} 
\end{eqnarray} 
where $c_{\gamma}=1,\, c_Z=R$ with  $R\equiv M_Z^2/M_W^2$.  
In Eqs. ($\ref{fer1}$) and ($\ref{fer2}$) 
\begin{eqnarray} 
L_f^2 \equiv  t^2(1-{4Q^2 \over M_W^2}\alpha(1-\alpha))+t(R_f-R_{f'}-1) 
+R_{f'}+i\epsilon \quad , 
\nonumber\ 
\end{eqnarray} 
and $C_g$ is the color factor (1 for leptons, 3 for quarks). 
Throughout this paper  $R_{i} \equiv {m_{i}^2 \over M_W^2}$
where the index $i$ refers to the relevant particle in each case.
The couplings 
$g_{L,R}^f$ appearing in Eqs. ($\ref{fer1}$) and ($\ref{fer2}$)    
are  $g_L^f=g_R^f=Q^f_{em}$  for $ V=\gamma $, and 
$g_L^f=Q^f_{wL}, g_R^f=Q^f_{wR}$ for $V=Z$.
% 
%\begin{eqnarray} 
%V=\gamma \quad & : & \quad\quad g_L^f=g_R^f=Q^f_{em} \nonumber\\ 
%V=Z      \quad & : & \quad\quad g_L^f=Q^f_{wL} \, , \, 
%                g_R^f=Q^f_{wR}  \nonumber\  
%\end{eqnarray} 
% 
 $Q^f_{em}$ are the electromagnetic charges, and $Q^f_{wL,R}$ are the 
weak charges for left/right handed fermions defined by the relation 
$Q^f_w\equiv T_3^f-Q^f_{em}\sin ^2 \theta _W$. 
$T_3^f$ is the weak isospin of the fermion $f$, ie. 
$T_3^f=-1/2$ for the left handed charged leptons etc.  
\par 
Regarding the contributions of the gauge bosons to $\Delta k_V$, $\Delta Q_V$  
,$(Q^{2}\neq 0)$, one faces the problem of gauge dependence.   
This is reflected in the loss of the  
crucial properties of infrared finiteness and perturbative unitarity.  
Analytic calculations of  
$\Delta k_V$, $\Delta Q_V$ were carried out in ref. ${\cite{argy}}$  
in the  't Hooft -- Feynman gauge. In that reference  
it was shown that $\Delta k_V$ has bad high  
energy behaviour, growing logarithmically as the energy increases, violating  
therefore unitarity constraints. At the same time $\Delta k_V$ exhibits an   
infrared ($IR$) singularity the only one surviving among several others   
that cancel each other. 
 
\par 
The problem of gauge dependence is simply related to the fact that in general 
$n$-point functions (propagators, vertices, etc) are not gauge invariant 
objects; only the $S$-matix elements have this property, being physically 
measurable quantities.   
Nevertheless there are several ways to extract the physically relevant 
information from their calculation. The most trivial example in our case, is to take the  
limit $Q^2\to 0$, which defines the static properties of $W$ boson. 
Less trivial is to consider special combinations of $\Delta k_\gamma$ 
and $\Delta k_Z$ which are directly measurable in the experiment. This is 
for instance the case of $e_L^+ e_R^- \to W^+W^-$ reaction, where the 
combination $\Delta k_\gamma - 4 Q^2/(4 Q^2-m_Z^2)\Delta k_Z$ can be measured. 
As is rather evident this combination is free of IR and gauge-dependence 
problems, as can be easily verified from the calculation of $\Delta k_V$ 
\cite{argy}.  
%A similar construction exists for $\Delta Q_V$. 
 
\par 
Of course, the gauge-dependence cancels out when the full one-loop matrix element 
for a specific process, for instance $e^+ e^- \to W^+W^-$, is performed. 
This means that the gauge-dependence of the three-point functions is eliminated 
from the corresponding contributions in the four-point functions.  
Therefore one can project out from the four-point functions the relevant 
tensor structures multiplying $\Delta k_V$ and $\Delta Q_V$ and after collecting 
all possible contributions to the specific process a gauge invariant 
and physically relevant result would be obtained.  
Nevertheless this procedure would require the knowledge of the full one-loop 
contributions for each specific process. On the other hand 
one can proceed in this direction by employing special field  
theory methods such as the Pinch Technique ($PT$) in order  
to define gauge-invariant and process-independent quantities.  
The $PT$ has been introduced by Cornwall  
$^{\cite{corn1}}$ and has since been employed in various physical   
problems ( see ref. ${\cite{corn1}} - {\cite{wat}}$). 
The main feature of the $PT$ algorithm, is that it enables us to  
define gauge-parameter independent self energies and vertices which  
satisfy QED-like Ward identities. More recent studies   
have explicitly demonstrated the universality of the $PT$-algorithm for gauge boson two   
point functions $^{\cite{wat}}$.  
Another approach, through which gauge-parameter{\footnote{With respect to 
gauge transformations of the background field.}} and process  
independent quantities can also be obtained, is the Background Field 
Method ($BFM$) $^{\cite{bfm}}$. In principle, $BFM$ provides a    
complementary and natural way to explain why the $PT$ rules yield vertex  
functions that have the desirable properties and satisfy simple Ward identities.  
  
\par  
More specifically in our case  
$PT$ provides us with an algorithm to isolate from box  
graphs pieces that have  
a vertex like structure and allot them to the usual vertex graphs. The   
details of this proceedure and the way one extracts the relevant pinch parts  
$ {\Delta {\hat k}}_{\gamma,Z}$  
can be traced in ref. ${\cite{papa}}$.  
Once the pinch contribution   
are taken into account the gauge boson contributions to TGV's   
become gauge independent  
approaching zero values as $Q^2$ increases, as demanded by unitarity, being  
also free of infrared singularities.    
  
\par 
Furthermore, by an explicit calculation, it can be shown that the pinch parts do  
not contribute when the incoming electron carries right handed chirality, as 
expected.  
In fact in that case the particular combination entering the amplitude  
for the process $e_L^+ e_R^- \to W^+W^-$ has vanishing   
pinch part and  hence in such a process only the genuine vertex parts  
contribute\cite{argy}. 
 
\par 
In the sequel and in order to keep our discussion independent of the   
specific process as much as possible we shall consider the $SM$   
contributions with the pinch parts properly incorporated into the  
usual vertex contributions.  
As yet, there is no systematic study of how well the $PT$  
approximates the coefficient of the relevant tensor structures of the full  
one-loop result in specific processes.  
Nevertheless, we shall use it as a basis for comparison   
of the supersymmetric contributions which are well behaved and not  
plagued by gauge dependent or infrared pathologies.  
%%%%%%%%%%%%%% 
%\vspace{.5cm} 
\vspace{.25mm} 
%%%%%%%%%%%%%%%%%%%%%%%%%%%%%%% section 2 %%%%%%%%%%%%%%%%%%%%%%%%%%%%%% 
\section*{2. The MSSM contribution to  
$\Delta k_{V}(Q^{2})$, $\Delta Q_{V}(Q^{2})$} 

The MSSM is described by a Lagrangian $^{\cite{Nilles}}$ 
\begin{equation} 
{\cal L}={\cal L}_{{SUSY}}+{\cal L}_{{soft}} 
\end{equation} 
where  ${\cal L}_{{SUSY}}$  is its  supersymmetric part and 
${\cal L}_{{soft}}$  the SUSY  breaking part. 
A convenient set of parameters to describe low energy physics 
is given by  
\begin{equation} 
m_{0} \quad,\quad A_{0} \quad,\quad M_{1/2} \quad,\quad 
\tan\beta(M_{Z}) \quad,\quad m_{t}(M_{Z}), 
\end{equation} 
where $m_{0}, A_{0}, M_{1/2}$ are the soft SUSY breaking parameters,
$m_{t}(M_{Z})$ is the value of the ``runing'' top quark mass at the 
Z - boson pole mass $M_{Z}$, and   $\tan\beta(M_{Z})$ the ratio of the 
v.e.v's of the Higgses  $H_2, H_1$. 
Complete expressions for the RGE's of all  
parameters involved can be found in the literature and will not be repeated  
here $^{\cite{Nilles}}$.   
\par 
At $Q^2=0$ the MSSM contributions to $\Delta k_{\gamma}$, $\Delta Q_{\gamma}$    
have been studied elsewhere and their dependences on the soft breaking 
parameters $A$, $m_0$, $M_{1/2}$, $\tan \beta$ and top quark mass $m_t$ 
have been investigated $^{\cite{spa}}$. 
In summary the conclusions reached  
in that paper are as follows : \\ 
i) Squarks and Sleptons have negligible effect on the dipole and quadrupole 
moments as compared to contributions of other sectors.\\ 
ii) The bulk of the MSSM Higgs contributions to $\Delta k_{\gamma}$, 
$\Delta Q_{\gamma}$  is due to the lightest CP - even neutral $h_0$, whose 
one loop corrected mass does not exceed $\approx 140 GeV$.    \\
iii) The Neutralinos and Charginos under certain conditions may be 
the dominant source of substantial deviations 
from the SM predictions. This happens for values of the soft gaugino mass 
$M_{1/2} \ll A_0, m_0$. 
In some extreme cases their contributions 
can even  saturate the LEP200 sensitivity limits.  
\par
The question of whether sizeable 
deviations from the SM predictions can show up at nonzero values of $Q^2$,  
accessible at LEP200 or other 
future colliders, is our principal motivation to study the behaviour of  
aforementioned quantities in the $Q^2 \not= 0$ region.

In the MSSM gauge boson and ordinary fermion contributions are like those 
of the SM and will be treated in the way desrcibed in the previous section. 
The additional  contributions from  
$\tilde{q}$, $\tilde{l}$  
(squarks, sleptons), $\tilde{Z}$, $\tilde{C}$ 
(neutralinos, charginos) as well as the supersymmetric Higgs contributions 
to $\Delta k_{V}$, $\Delta Q_{V}$ can be deduced from the triangle  
graph shown in Figure 1 where $P_{1,2,3}$ are the appropriate internal  
lines. We will not consider graphs  
that yield vanishing contributions to both 
$\Delta k_{V}$ and $\Delta Q_{V}$.  
 
In the following we shall consider   
the contributions of each sector separately. 
\subsection*{Squarks--Sleptons ($\tilde{q}$, $\tilde{l}$)}  
\par 
We first consider the contributions of the sfermion sector of the theory  
which can be read from the diagram of Figure 1 with the assignment 
$P_1=P_2=\tilde f, P_3={\tilde f}^\prime $ to the internal lines. 
 $ \tilde f$ and ${\tilde f}^\prime $ denote sfermions. 
Unlike matter fermions  
this graph involves mixing matrices due to the fact that left  ${\tilde f}_L$  
and right  ${\tilde f}_L^c$  handed sfermions mix when electroweak symmetry  
breaks down. Such mixings are substantial in the stops, due to  
the heaviness of the top quark, resulting in large mass splittings of the  
corresponding mass eigenstates ${\tilde t}_{1,2}$. \\ 
In units of $g^2/{16\pi^{2}}$  
the sfermion contributions are given by: 
%\newpage 
% 
%%%%%%%%%%%%%%%%%%%%%%%%%%%%%%%%%%%%%%%%%%%%%%% 
% 
\begin{eqnarray} 
 \Delta k_V &=&      
      - 2 C_g c_V T_3^{ f}  \sum_{i,j,k=1}^{2}     
 ({  A_{ij}^{V,{\tilde f} } } {K_{i1}^{\tilde f} }{K_{j1}^{\tilde f}})                
     ({K_{k1}^{\tilde f'} } {K_{k1}^{\tilde f'} })  \nonumber  \\ 
&  &  {\times} \quad   \olocl {  
  t^2 (1-t)[2t-1+R_{{\tilde f}_i} \al + R_{{\tilde f}_j} (1- \al) 
 - R_{{\tilde f'}_k}] \over 
	    L_{{\tilde f}{\tilde f'}} ^2  }   \\  
     & &   \nonumber            \\ 
 \Delta Q_V &=& 
      8 C_g c_V T_3^{ f}  \sum_{i,j,k=1}^{2}     
 ({  A_{ij}^{V,{\tilde f} } } {K_{i1}^{\tilde f} }{K_{j1}^{\tilde f}})                
     ({K_{k1}^{\tilde f'} } {K_{k1}^{\tilde f'} })  \nonumber   \\ 
& &  {\times} \quad    \olocl { { t^3(1-t)\al (1-\al)}  \over 
      L_{{\tilde f}{\tilde f'}} ^2 }     
\end{eqnarray}         
where 
\begin{eqnarray}    
 L^2_{\tilde f,\tilde f'} & \equiv  & 
 t^2+(R_{\tilde f_{i}} \al +  R_{\tilde f_{j}} (1-\al) -1) t   
 +R_{\tilde f^{\prime }_{k}}(1-t) -\usual .  \nonumber      
% \\ 
% (R _{ {\tilde f}_i , {{\tilde  f}^\prime}_i }   
%    & \equiv &  {(m_ { {\tilde f}_i , {{\tilde  f}^\prime}_i }  
%    / M_W)}^2) \quad .    \nonumber  
\end{eqnarray} 
In the expressions above $A_{ij}^{V,{\tilde f}}$ are given by,
%\begin{eqnarray}    
$  {  A_{ij}^{Z,{\tilde f} } } = Q_W^{\tilde f} {{\delta}_{ij}} - 
  T_3^{\tilde f} {K_{i2}^{\tilde f} } {K_{j2}^{\tilde f}} \quad, \quad 
  {  A_{ij}^{\gamma,{\tilde f}}}  
  = Q_{em}^{\tilde f} {{\delta}_{ij}} $  %  \qquad  , \label{AAA} 
%\end{eqnarray} 
%%%%%%%%%%%%%%%%%%%%%%%%%%%%%%%%%%%%%%%%%%%%%% 
% 
where $Q^{\tilde f}_{W}$ and     
$Q^{\tilde f}_{W} \equiv T_3^{\tilde f}-Q^{\tilde f}_{em}\sin ^2 \theta _W$ 
are the electromagnetic and weak charges of the sfermion ${\tilde f}$
respectively.
The remaining factors  
$c_{\gamma},\, c_Z $ , and 
$C_g$  were defined in the previous section.  
${\tilde f}_{1,2}$ and ${\tilde f'}_{1,2}$  
denote the mass eigenstates while ${\bf K}^{{\tilde f},{\tilde f'}}$ 
diagonalize 
the corresponding mass matrices (for notation see  ref.\cite{spa}).
%%%%%%%%%%%%%%%%%%%%%%%%%%%%%%%%%%%%%%%%%%%%%% 
%${\bf K}^{{\tilde f},{\tilde f'}}  
%{\bf {\cal M}}^{ 2}_{{\tilde f},{\tilde f'}} 
%{{\bf K}^{{\tilde f},{\tilde f'}}}^{\bf T} =diagonal$.  
%%%%%%%%%%%%%%%%%%%%%%%%%%%%%%%%%%%%%%%%%%%%%% 
In the stop sector for instance,  
the diagonalizing matrix is defined as 
${\bf K}^{\tilde t} {\bf {\cal M}}^{ 2}_{ {\tilde t}}  
{{\bf K}^{\tilde t}}^{\bf T}  
=diagonal(m^2_{{\tilde t}_1},m^2_{{\tilde t}_2})$. 
In the absence of \, SUSY breaking effects $m_{{\tilde f},{\tilde f'}} = 
m_{f,f'}$ and ${\bf K}^{{\tilde f},{\tilde f'}}$ become the unit matrices. In   
that limit $\Delta Q_V$ given above cancels against the corresponding  
fermionic contribution of Eq. ($\ref{fer2}$) as it should. 
%\vspace{.6cm} 
\vspace{.25mm}  
%%%%%%%%%%%%%% 
\subsection*{Neutralinos--Charginos ($\tilde{Z}$, $\tilde{C}$)}          
The neutralino and chargino sector is perhaps the most awkward sector to deal 
with due to mixings originating from the electroweak symmetry breaking  
effects.  
The two charginos $\tilde{C}_i$ (Dirac fermions) and the four 
neutralinos $\tilde{Z}_\alpha$ (Majorana fermions) 
are eigenstates of the  mass matrices ${\bf {\cal M}_{\tilde{C}}} $ and
  ${  \bf {\cal M}_{\tilde{N}}} $ whose explicit expressions 
can be found in ref. \cite{spa}.     
If the ${\bf U}$, 
${\bf V}$ matrices diagonalize ${\bf {\cal M}_{\tilde{C}}}$, i.e. 
${\bf U{\cal M}_{\tilde{C}} V^{\dagger}}=diagonal$, and ${\bf O}$ 
(real orthogonal) diagonalizes the real symmetric neutralino mass matrix
${  \bf {\cal M}_{\tilde{N}}} $,  
 ${\bf O^{T}{\cal M}_{\tilde{N}} O}=diagonal$, then their  
electromagnetic and weak currents are given by 
%\newpage 
% 
\begin{eqnarray} 
{J_{em}^\mu}&=&{\sum_{i}} {\bar{\tilde C}}_i  \gamma^\mu {\tilde C}_i \quad , 
\quad {J^\mu_+}={ \sum_{\alpha,i}}{\bar {\tilde Z}}_\alpha \gamma^\mu  
      ( P_R C^R_{\alpha i}+P_L C^L_{\alpha i}) {\tilde C}_i  \nonumber \\ 
{J^\mu_0}&=&{ \sum_{i,j}}{\bar {\tilde C}}_i \gamma^\mu  
      ( P_R A^R_{ij}+P_L A^L_{ij}) {\tilde C}_j+  
 {1 \over 2} { \sum_{\alpha,\beta}}{\bar {\tilde Z}}_\alpha \gamma^\mu  
      ( P_R B^R_{\alpha \beta}+P_L B^L_{\alpha \beta}) {\tilde Z}_\beta  
  \nonumber    
\end{eqnarray} 
where  
\begin{eqnarray} 
C^R_{\alpha i}&=&-{\frac {1}{\sqrt 2}} O_{3 \alpha} U^{*}_{i2}- 
				     O_{2 \alpha} U^{*}_{i1} \;\;,\;\; 
C^L_{\alpha i}=+{\frac {1}{\sqrt 2}} O_{4 \alpha} V^{*}_{i2}- 
				     O_{2 \alpha} V^{*}_{i1} \\      
A_{ij}^h&=&[\cos^2\theta_W \delta_{ij}-{1 \over 2}(V_{i2}V_{j2}^* \delta_{hL}                                      
       +U_{i2}^*U_{j2} \delta_{hR})]\;\;,\;\;h=L,R \\ 
B_{\alpha \beta}^L&=&{1 \over 2}(O_{3 \alpha}O_{3\beta}-O_{4 \alpha}O_{4\beta}) 
      \quad,\quad B_{\alpha \beta}^R=-B_{\alpha \beta}^L\:.  
\end{eqnarray} 
$P_{R,L}$ are the right/left$-$handed projection operators $(1\pm \gamma_5)/2$. 
 
The contributions of this sector to $\Delta k_{\gamma}$, $\Delta Q_{\gamma}$, 
 which are given below, 
are calculated from the graph of Figure 1,  
with the following assignments to the internal   
lines $P_{1,2,3}$, \\  
 
%\vspace*{.2cm} 
%Graph of Figure 1 with  
{\underline {$(P_1,P_2,P_3)=(\tilde C,\tilde C,\tilde Z)$}} : 
\begin{eqnarray} 
\Delta k_{\gamma}&=&-\sum_{i, \al}\olocl \{ F_{\al i}  
	  [t^4+(R_{\al}-R_i-1)t^3+(2R_i-R_{\al})t^2    \nonumber\\ 
     & & + \usualp ] + G_{\al i} {{m_i m_\al}\over {M_W^2}}(4t^2-2t) \}            
     {1 \over L_{\tilde Z}^2} \\ 
\Delta Q_{\gamma}&=&-8\sum_{i ,\al} F_{\al i} \olocl {t^3(1-t)\al  
	  (1-\al) \over  L_{\tilde Z}^2},  
\end{eqnarray} 
where \hspace{.7cm} 
$F_{\al i}=\mid C_{\al i}^R \mid ^2+ \mid C_{\al i}^L \mid ^2 \quad, 
  \quad G_{\al i}\;=\;(C_{\al i}^L C_{\al i}^{R*}+h.c.) $, and
\begin{eqnarray} 
L_{\tilde Z}^2&=&t^2+(R_i-R_{\al}-1)t+R_{\al}-\usual . 
\end{eqnarray} 
In the equations above
the indices $\al=1,2,3,4$ and $i=1,2$ refer to neutralino and  
chargino states respectively. Note that we have not committed 
ourselves to a particular  
sign convention for the masses $m_i$, $m_{\al}$ appearing in the sum in 
the equation above for the $\Delta k_{\gamma}$. 
Chiral rotations that make these masses positive also affect 
the rotation matrices and should be taken into account. 
\par 
%\newpage 
For the couplings $\Delta k_Z$, $\Delta Q_Z$ we have the following 
contributions\\ 
 
%Graph of  Figure 1 with  
i) \hspace{5mm} 
{\underline {$(P_1,P_2,P_3)=(\tilde C,\tilde C,\tilde Z)$}} : 
\begin{eqnarray} 
\Delta k_Z&=&-R\;\sum_{i,j,\al}\olocl \{ S^L_{ij\al} [t^2(1-t)(-t 
     +R_i\al +R_j (1-\al)-R_{\al})\nonumber\\ 
   & &+\usualp]+ \,{m_im_j \over M_W^2} S^R_{ij\al} t^2 \nonumber \\ 
 & &-{m_im_{\al} \over M_W^2} (T^L_{ij\al}+T^R_{ij\al})[2t^2\al +t^2-t] \} 
 {1 \over L^2_{ij\al}} \\ 
\Delta Q_Z&=&-8R\;\sum_{i,j,\al} 
		  S^L_{ij\al}\olocl{t^3(1-t)\al (1-\al) \over L^2_{ij\al}} 
\end{eqnarray} 
where  
\begin{eqnarray} 
S^{L(R)}_{ij\al}&\equiv&(C^{L*}_{\al i}C^{L}_{\al j}A^{L(R)}_{ji} 
		     +(L\rightleftharpoons R))\quad,\quad 
T^{L(R)}_{ij\al}\equiv(C^{L*}_{\al i}C^{R}_{\al j}A^{L(R)}_{ji} 
		     +(L\rightleftharpoons R)) \nonumber \\ 
L^2_{ij\al}&=&\al tR_i+t(1-\al)R_j+R_{\al}(1-t)-t(1-t)-\usual \nonumber 
\end{eqnarray} 
The indices $i,j$ refer to charginos 
and $\al $ to neutralino mass eigenstates.
\vspace{.2cm} 
\\ 

\noindent
%Graph of  Figure 1 with  
ii) \hspace{5mm}   
{\underline {$(P_1,P_2,P_3)=(\tilde Z,\tilde Z,\tilde C)$}} : \\  \\  
%\vspace*{.2cm}  
%% 
This is the same as the 
previous graph with $\{ i, j, \alpha \}$ replaced by 
$\{ \rho, \sigma, i \}$  and 
$ S^{L(R)}_{ij\rho}, T^{L(R)}_{ij\rho}$, 
$L^2_{ij\al}$ replaced by the following expressions: \\ 
\begin{eqnarray}    
S'^{L(R)}_{\rho \sigma i} &\equiv& -(C^{L*}_{\rho i}C^{L}_{\sigma i} 
	    B^{L(R)}_{\rho \sigma}+(L\rightleftharpoons R)) \quad , \quad
T'^{L(R)}_{\rho \sigma i} \equiv -(C^{R*}_{\rho i}C^{L}_{\sigma i} 
	    B^{L(R)}_{\rho \sigma}+(L\rightleftharpoons R))   \nonumber \\ 
L'^2_{\rho \sigma i}&=&\al tR_{\rho}+t(1-\al)R_{\sigma} 
	 +R_i(1-t)-t(1-t)-\usual \nonumber 
\end{eqnarray} 
In these ${\sigma,\rho}$ refer to neutralinos and $i$ to chargino mass 
eigenstates.
%% 
%\vspace{.6cm} 
\vspace{.25mm}   
%%%% 
\subsection*{Higgses ($H_0,h_0,A,H_{\pm}$)} 
%%%%%%% 
There are five physical Higgs bosons which survive electroweak symmetry 
breaking. Two of these, $H_0$ and $h_0$, are neutral and $CP$ even, while 
a third $A$, is neutral and $CP$ odd. The remaining Higgs bosons, 
$H_{\pm}$, are charged. At the tree level the lightest of these, namely 
$h_0$, is lighter than the $Z$ gauge boson itself. However 
it is well known that radiative corrections, which are 
due to the heavy top, are quite large and 
may push its mass above $M_Z$. 
$h_0$ turns out to yield the largest contributions of all Higgses 
to the TGV's, since the remaining Higgses have 
large masses of the order of the SUSY breaking scale. 
At the tree level the masses of all Higgs bosons involved are given by the  
following expressions : 
\begin{eqnarray} 
{m^2_A}&=& -{2 B \mu \over \sin2\beta}\quad,\quad  
 m^2_{H_\pm}=m^2_A+ M_W^2  \\
{m_{ H_0,h_0 }^2}&=&{\frac {1}{2}} \{  {(m^2_A+M_Z^2)}^2\pm      
  \sqrt {   {(m^2_A+M_Z^2)}^2 -4{M_Z^2} {m^2_A} {cos^2}(2\beta) }   
								 \}   
\end{eqnarray} 
The Higgs contributions can 
be expressed in terms of their masses and an angle $\theta$, which relates  
the states $S_1 \equiv \cos \beta \: (Real \; H_1^0) + 
\sin \beta \: (Real \; H_2^0)$\, , \,  
$S_2 \equiv -\sin \beta \: (Real \; H_1^0) + 
\cos \beta \: (Real \; H_2^0)$ to the mass eigenstates $h_0,H_0$.  
The state $S_1$ is the SM Higgs boson, which however is not a mass  
eigenstate since it mixes with $S_2$.  
When $\sin ^2 \theta =1$ such a mixing does not occur and $h_0$ becomes  
the standard model Higgs boson $S_1$. 
\par 
The various contributions of the Higgs bosons to  
$\Delta k_{\gamma}$, $\Delta Q_{\gamma}$, are given below. The assignment   
to the internal lines $(P_1,P_2,P_3)$ in each case is explicitly shown: 
%%%% 
%\vspace{.2cm} 
%%%% 
%%%%%%%%%%%%%%%%%%%%%%%%%%%%% 
\begin{eqnarray} 
&& i) \hspace{10mm} 
 {\underline { (P_1,P_2,P_3) = (H_+,H_-,A)}} : \hspace{85mm} \nonumber \\ 
&&\hspace{20mm}  \Delta k_{\gamma}=D_2(R_A,R_{H_{+}})  \quad     , \quad 
	    \Delta Q_{\gamma}=Q(R_A,R_{H_{+}}) \\ 
&& ii) \hspace{10mm} 
{\underline{ (P_1,P_2,P_3) = (W_+,W_-,h_0)+(H_+,H_-,h_0)}} :  
 \nonumber \\  
&&\hspace{20mm} \Delta k_{\gamma}\: = \: \sin^2 \theta \: D_1\, 
			 (R_{h_0})\:+\:\cos^2 \theta \:  
			 D_2(R_{h_0},R_{H_{+}})  \label{pr1} \\ 
&&\hspace{20mm} \Delta Q_{\gamma}\:=\: \sin^2 \theta \: 
			     Q(R_{h_0},1)\:+\: \cos^2 \theta  
				  \:Q(R_{h_0},R_{H_{+}}) \label{pr2}   \\ 
&& iii) \hspace{10mm}   
{\underline{ (P_1,P_2,P_3) = (W_+,W_-,H_0)+(H_+,H_-,H_0)}} :  
    \nonumber  \\ 
&&\hspace{20mm} As \quad in \; ii) \quad with  
	      \quad R_{h_0} \rightarrow R_{H_0} 
\quad and 
       \quad \sin^2 \theta \rightleftharpoons  \cos^2 \theta    
\end{eqnarray} 
where 
  $  \sin ^2\theta=({M_A^2+M_Z^2\sin ^2 2\beta -M_{h_0}^2})  /
	  ( { M_{H_0}^2-M_{h_0}^2}) \:  $.
The functions $D_{1,2}, Q$ appearing above are given by 
\begin{eqnarray} 
&&   D_1(r)\equiv{1\over 2} \olocl 
	{ 2t^4+(-2-r)t^3+(4+r)\,t^2   \over 
	t^2+r(1-t)-\usual }      \\ 
&&    D_2(r,R)\equiv{1\over 2} \olocl 
	{ 2t^4+(-3-r+R)t^3+(1+r-R)t^2   \over 
	t^2+(-1-r+R)t+r-\usual }  \\ 
&&    Q(r,R)\equiv 2 \olocl 
	{ t^3(1-t)\al (1-\al) \over 
	 t^2+(-1-r+R)t+r-\usual }   
\end{eqnarray} 
\par 
For the $\Delta k_Z$, $\Delta Q_Z$ form factors we get 
the following contributions:
%\\ 
%%%% 
\\ \\ 
%\newpage 
i) \hspace{5mm}    \underline {$(P_1,P_2,P_3)=(W_+,W_-,h_0+H_0)$} : \\ 
\begin{eqnarray} 
\Delta k_{Z}&=&{1\over 4}\{ (\cos^2 \theta) \olocl\; [(4-2R)t^4+(R-2) 
     (R_{H_0}+2)t^3 \nonumber\\ 
   & &+(2R_{H_0}-RR_{H_0}+8-2R)t^2] {1 \over L^2_{H_0}} 
     + (\sin^2 \theta)\times (H_0 \rightarrow h_0) \} \\ 
\Delta Q_{Z}&=&(2-R) \{ (\cos^2 \theta) \olocl {t^3(1-t) \al (1-\al) 
   \over L^2_{H_0} } \nonumber\\ 
 & & +(\sin^2 \theta)\times (H_0 \rightarrow h_0) \} 
\end{eqnarray} 
where \hspace{4mm}
$  L^2_{H_0} \equiv t^2+R_{H_0}(1-t)-\usual $.    \\  \\ 
%%% 
%\vspace*{.4cm} 
ii) \hspace{5mm}    \underline {$(P_1,P_2,P_3)=(H_+,H_-,A+H_0+h_0)$}: \\  
\begin{eqnarray} 
\Delta k_{Z}=({2-R \over 2}) \{ D_2(R_A,R_{H_{+}})+\sin^2 \theta  
     \;D_2(R_{H_0},R_{H_{+}})+\cos^2 \theta \;D_2(R_{h_0},R_{H_{+}}) \} \\ 
\Delta Q_{Z}=({2-R \over 2}) \{ Q(R_A,R_{H_{+}})+\sin^2 \theta\;  
    Q(R_{H_0},R_{H_{+}}) +\cos^2 \theta \;\;Q(R_{h_0},R_{H_{+}}) \}  
\end{eqnarray} 
%%% 
%\vspace{.5cm} 
\\ 
iii) \hspace{5mm}  
\underline {$(P_1,P_2,P_3)=(H_0+h_0,A,H_+)+(A,H_0+h_0,H_+)$}: \\  
\begin{eqnarray} 
\Delta k_{Z}&=&{R\over 2} \{ (\sin^2 \theta) \olocl\; {t^2(1-t)(1+R_{H_{+}} 
     -R_A\al -R_{H_0}(1-\al)-2t) \over {\tilde L}^2_{H_0}} \nonumber\\ 
& & \hspace{7.2cm} + (\cos^2 \theta)\times (H_0 \rightarrow h_0) \} \\ 
\Delta Q_{Z}&=&2R \{ (\sin^2 \theta) \olocl {\al (1-\al) t^3(1-t) 
   \over {\tilde L}^2_{H_0} } 
    +(\cos^2 \theta)\times (H_0 \rightarrow h_0) \} \\ 
     & & \nonumber 
\end{eqnarray} 
with \hspace{2mm}
${\tilde L}^2_{H_0} \equiv -t(1-t)+R_A\al t+R_{H_0}(1-\al)t 
		   + R_{H_{+}}(1-t)-\usual $  ,    
\newline
and finally,\\ \\ 
iv) \hspace{5mm} 
\underline {$(P_1,P_2,P_3)=(Z,H_0+h_0,W_+)+(H_0+h_0,Z,W_+)$}: \\     
\begin{eqnarray} 
\Delta k_{Z}&=&{R\over 2}\{ (\cos^2 \theta) \olocl\;  
    [-6\al t^2+(t^3-t^2)  (2(t-1)+(R-R_{H_0})\al +R_{H_0}) \nonumber\\ 
   & &+2(R-1)\al t^2] {1 \over {\hat L}^2_{H_0}} 
     + (\sin^2 \theta)\times (H_0 \rightarrow h_0) \} \\ 
\Delta Q_{Z}&=&2R \{ (\cos^2 \theta) \olocl {\al (1-\al)t^3 (1-t) 
   \over {\hat L}^2_{H_0} } \}  
    +(\sin^2 \theta)\times (H_0 \rightarrow h_0) \} 
\end{eqnarray} 
with
${\hat L}^2_{H_0} \equiv  (1-t)^2+R\al t+R_{H_0}t(1-\al)-\usual  $  . 
%%% 
\\  \\
In most of the parameter space 
the Higgses $A, H_{\pm}$ and $H_0$ turn out to be rather heavy with 
masses of the order of the SUSY breaking scale; therefore all graphs in  
which at least  
one of these participates are small. At the same time $\sin^2 \theta$ has a 
value very close to unity. Thus the dominant Higgs contribution arises solely 
from graphs in which a $h_0$ is exchanged. This 
is exactly what one gets in the SM with $h_0$ playing the role of the SM 
Higgs boson. \\ 
 
The form factors considered so far develop also imaginary (absortive) 
parts which show up as we cross the thresholds of internal particles in  
the loop. These can be calculated using the $i \epsilon$ prescription. 
Note also that the pinch parts discussed before have imaginary 
parts which should be taken into account as we do.  
The absortive parts contribute 
to physical processes too and therefore for a complete analysis their 
behaviour as a function of the energy should be studied. For lack of  
space in this paper we do not display analytic expressions for the  
imaginary parts. Their behaviour as a function of the variable $Q^2$ 
will be discussed in our conclusion part. 
% 
%%%%%%%%%%%%%%%%%%%%%%%%%%%%%%% Section 3 %%%%%%%%%%%%%%%%%%%%%%%%%%%%%% 
%\vspace{.6cm} 
\vspace{.25mm}   
%%%%%%%%%% 
\section*{3. Numerical Analysis -- Conclusions } 
 
Both dispersive and absortive parts of  
the trilinear $WW\gamma$, $WWZ$ vertices can be cast in the form of
single integrals of  
known functions of $t$ and $Q^2$, which also depend on the physical masses  
of all particles involved. 
In fact wherever double $\olocl$ integrations are encountered we first  
perform the $\int_{0}^{1} d\alpha$ integrations explicitly and subsequently  
carry out the $\int_{0}^{1} dt$ integrations numerically  using special     
routines of the  FORTRAN Library $IMSL$ .  
The advantage of using  
this facility is that it leads to reliable results even in cases where the  
integrands exhibit fast growth at some points or have a rapid oscillatory  
behaviour. The inputs in these calculations are  
the variable $Q^2$ and the arbitrary parameters of the 
MSSM. 
\par 
With the experimental inputs $M_Z=91.188 \: GeV$, $\sin^{2} {\theta_W} =.232$, 
$\alpha_{em}(M_Z)=1/127.9$ and  $\alpha_{s}(M_Z)=.115$ and with given values 
for the arbitrary parameters  $\tan \beta (M_Z)$, $m_t (M_Z)$,  
$A_0$, $m_0$, $M_{1/2}$ 
we ran our numerical routines in order to know the mass spectrum and the  
relevant mixing parameters necessary for the evaluation of the form factors  
given in the  
previous sections.
For the running top quark mass $m_t(M_{ Z})$   we took values  
in the whole range from $130\: GeV$ to $190\: GeV$,  
although small values of $m_t$ are already ruled out     
in view of the $CDF$ and $D0$ results $^{\cite{abe}}$.     
The physical top quark masses emerging are slightly larger by  
about $3 \% $         

As for the soft SUSY breaking parameters  
$A_0,m_0,M_{1/2}$ we scan the three dimensional parameter space from  
$\simeq 100\: GeV$ to $1\: TeV$.  
This parameter space can be divided into three 
main regions 
i) $A_0 \simeq m_0 \simeq M_{1/2} $ (SUSY breaking terms comparable),  
ii) $A_0 \simeq m_0 \ll M_{1/2} $ (the gaugino mass is the dominant source  
of SUSY breaking ) and 
iii) $M_{1/2} \ll  A_0 \simeq m_0 $ ($A_0, m_0$ dominate over $M_{1/2}$) .
Case ii) covers the no-scale models for which the  
preferable values are $A_0=m_0=0$, while case iii) covers the light  
gluino case. 
\par 
Regarding the values scanned for the energy variable $Q^2$ we explored both  
the timelike and spacelike regions for values ranging from  
$\mid Q^2 \mid \, = \, 0$ 
to ${ \mid Q^2 \mid} \,=\,10^5 {M_W^2}$. For the timelike case, which is of  
relevance for future collider experiments, this corresponds to values of  
$\sqrt s$ ranging from $0\: GeV$ to about $600\: M_W$.  
Both in the spacelike  
and timelike energy region we have seen that 
as soon as $\sqrt s$  exceeds $\simeq$ few $TeV$ 
the contributions of each sector separately becomes negligible, approaching  
zero as the energy increases in accord with unitarity requirements. 
\par 
Sample results are presented in Table 1 for values of  
$(A_0,m_0,M_{1/2})$ equal 
to  $(300,$ $300,$ $300)$, $(0,0,300)$ and $(300$, $300$, $80)$ $GeV$ 
representative of the cases i),ii) and iii)  discussed 
previously. The inputs for   
the remaining parameters are  
\mbox{$tan\beta (M_Z) = 2$}, $m_t(M_Z)=170\: GeV$. 
The value of $\sqrt s$ in these tables are respectively $190$ and $500\:GeV$, 
corresponding to the center of mass energies of LEP200 and NLC. 
\footnote{Throughout this paper whenever we refer 
to MSSM predictions we mean both sparticle and particle contributions,
gauge boson and ordinary fermion contributions inclusive.}
In the same 
table for comparison we give the SM predictions for Higgs masses 
equal to $50, 100$ and $300\: GeV$. With the inputs given above the typical 
SUSY breaking scale lies somewhere between $2 M_W$ and $0.5 \:TeV$.  
Although many 
sparticle thresholds exist in this region, as for instance the lightest  
of the sleptons and squarks as well as the lightest of the neutralinos  
and charginos, especially when $M_{1/2}$ is light, these thresholds do not 
result in any enhancement of the form factors 
$\Delta k_{\gamma,Z}$, $\Delta Q_{\gamma,Z}$. Increasing the value of the 
dominant SUSY breaking scale the supersymmetric contributions to these 
quantities become less important approaching zero values. 

We now come to discuss separately
the contributions of sparticles and supersymmetric Higgses.
The sfermions yield the smaller contributions 
even in cases where due to large electroweak mixings some of the squarks, 
namely one of the stops, are relatively light. The supersymmetric Higgses  
yield contributions comparable to those of the SM, provided the latter  
involves a light Higgs with mass around $100\:GeV$. The bulk of the  
Higgs contributions is due to the lightest $CP$ even neutral $h_0$.  
Thus SUSY Higgs contributions are like those of the SM
with $h_0$ playing the role 
of the Standard Model Higgs boson. 
The neutralinos and charginos in some cases,  
depending on the given inputs, can accomodate light states. Their  
contributions in that case are not necessarily suppressed and are the principal 
source of deviations from the SM predictions. The 
contributions of this sector are sensitive to the input 
value for the soft gaugino mass $M_{1/2}$, being more important for values 
$M_{1/2} < 100 \; GeV$. For such values of the soft gaugino mass and in the  
region, 
$0 < \sqrt{s} < 2 M_W $ their contributions are enhanced, due  
to the development of an anomalous threshold, 
resulting to sizeable corrections to the magnetic dipole and electric  
quadrupole moments of the $W$ boson $^{\cite{spa}}$. 
However such an enhancement does not occur in the physical region and these
contributions fall rapidly to zero as we depart from the unphysical region 
to values of energies above the two $W$ production threshold.  
\par 
%$\Delta k_{\gamma,Z}$, $\Delta Q_{\gamma,Z}$  
The total contributions to the $\Delta k_{\gamma,Z}$, 
for some particular cases,  
both in the MSSM   
and SM are shown in Figure 2 for  
values of $\sqrt{s}$   
ranging from $0\: GeV$ to $1\: TeV$. 
The region from $0$ to $2M_W$ is unphysical since the external $W$'s have  
been taken on their mass shell. At $s = 0$ the quantities  
$\Delta k_{\gamma}$, $\Delta Q_{\gamma}$ are linearly related to the magnetic 
moment and electric quadrupole moments of the $W$-boson. The MSSM 
predictions displayed in Figure 2 are for the particular choice 
$(A_0,m_0,M_{1/2})$ =  (300, 300, 80), the  
most interesting of the three cases discussed previously, since it  
includes light neutralino and chargino states. 
We only show the $\mu > 0$ case. The negative $\mu$ case leads 
to similar results. 
For lack of space we have only displayed the dispersive   
part.
The absortive part turns out to have qualitatively a similar behaviour 
and we will not discuss them any further. 
All form factors 
tend to zero fairly soon with increasing energy  
reaching their asymptotic values at energies  
\mbox{$\sqrt{s} \approx few \; TeV$} in    
agreement with unitarity constraints. 
The first peak observed in the unphysical region  
\mbox{( $ \sqrt{s} < 2 M_W $ )}   
in the MSSM is due to the anomalous threshold of the  
Neutralino/Chargino sector discussed previously.  
In the physical region \mbox{( $  \sqrt{s} > 2 M_W $) }  
the first peak observed  
is associated with the $t \bar{t}$ production threshold while the second,  
around $700\: GeV$, is due  
to the threshold of the heavy neutralino states. 
We should point out that if it were not for the   
Neutralino/Chargino sector the MSSM and SM predictions would differ little. 
This sector is therefore the dominant source of deviations from the  
Standard Model predictions provided it accomodates light states. 
\par 
The TGV's studied in this paper have been promoted to physical observables 
as being gauge independent satisfying at the same time the perturbative 
unitarity requirements, using the PT algorithm. 
The question of how to extract 
physical information on these 
vertices from measurements of physical processes at $e^+e^-$ and hadron 
colliders has 
been the subject of many phenomenological studies and has recently triggered 
numerous debates among physicists in working groups at LEP200 
\cite{lep200-tgc} and other 
workshops [D. Zeppenfeld in {\cite{flei}}, {\cite{aihara}}]. 
$WW\gamma$ and $WWZ$ vertices are parts of the experimentally observed 
amplitudes of the $e^+ e^- \rightarrow W^+ W^-$ process whose SM radiative 
corrections have been studied elsewhere {\cite{lemoine}}. 
The accuracy of measuring 
the aforementioned vertices is improved with polarized beams. 
%In this case particular combinations 
%of the form factors under consideration enter the helicity amplitudes 
%{\cite{gounar,flei2}}.
%enter the helicity amplitudes whose 
%square are measurable quantities {\cite{gounar,flei2}}.  
In this case, it is the particular combination  
$\Delta k_\gamma - 4 Q^2/(4 Q^2-m_Z^2)\Delta k_Z$  
which enters the amplitude for the reaction  
$e_L^+ e_R^- \to W^+W^-$ {\cite{gounar,flei2}}, and this is gauge independent 
and free of infrared singularities. Moreover it is independent of the 
PT algorithm, as we have already discused. Its behaviour as a function 
of the energy is displayed in Figure 3 for both MSSM and SM. The 
parameters are as in Figure 2.  
In the MSSM both $\mu > 0$ and $\mu < 0$ cases are shown. 
One notices that  
MSSM and SM predictions for this quantity are very close to one another 
apart from the peak which is due to the heavy neutralino threshold. 
Differences however are small, and to be detected requires high 
sensitivities. 
\par 
In our analysis 
we have focused our attention on the TGV's considering all SUSY  
contributions to them. For a complete one loop analysis of the  
$e^+ e^- \rightarrow W^+ W^-$ process the SUSY box contributions should be  
taken into account. There is no a priori reason why these should be small. 
However these graphs involve exchanges of at least one selectron or 
sneutrino whose masses are large of the order of the supersymmetry  
breaking scale. Our previous considerations on the TGV's has shown that 
sleptons yield smaller contributions as compared to other sectors and  
especially that of the neutralinos and charginos. On these grounds we expect 
a small effect from the supersymmetric box graphs in which sleptons 
are exchanged. 
Indeed it has been shown$^{\cite{kneur}}$ 
that the effect of some supersymmetric boxes
is quite small at LEP200 energies.
At NLC energies however these become comparable to the TGV contributions.
Therefore for a complete phenomenological study 
their contributions should be taken into account.
The effect of the box graphs
on the TGV form factors entering into $e^+ e^- \rightarrow W^+ W^-$ 
is under study and the results will appear in a future publication.   
\par 
Our conclusion concerning the trilinear gauge boson vertices is that for  
energies 
$2 M_W < \sqrt{s} < 200 \; GeV $ the MSSM predictions differ in general  
from those of the  
\mbox{SM} but they are of the same order of magnitude. To be detectable 
deviations from the SM 
requires sensitivities reaching the per mille  
level and hence  unlikely to be observed at LEP200. If deviations from the  
SM predictions are observed at these energies it will be a signal of new  
underlying dynamics, which however will not be of supersymmetric nature. 
At higher energies  SM and 
MSSM predictions fall rapidly to zero, due to unitarity,  
getting smaller, in most cases, 
by almost an order of magnitude already at energies  
$\sqrt{s} \approx 0.5  \; TeV $. As a result, the task of  
observing deviations from the SM which are due to supersymmetry  
demands higher experimental accuracies as well as a complete
theoretical treatment which properly takes care of box contributions.
\\  \\
%\vspace*{.5cm} 
\noindent 
%\newpage
{\bf {\large {\bf Acknowledgements}}}  
\\ 
%\vspace*{1mm} 
This work was supported by the EEC Human  
Capital and Mobility  
Program, CHRX -- CT93 - 0319. The work of A. B. L. was also  supported by 
the EEC Science Program SCI-CT92-0792.\\[24pt] 
%%%%%%%%%%%%%%%%%%%%%%%%%%%%%%%%%%%%%%%%%%%%%%%%% 
%\newpage 
% 
 
\newpage 
%\noindent 
%{\large {\bf Table Captions}} 
 
\vspace{.4cm} 
\begin{table} 
\begin{center}     
\vspace*{1.cm}      
%\caption{ } 
\begin{tabular}{|cccc|} \hline 
%\multicolumn{4}{|c|}{ \bf TABLE I} \\ 
%\hline  
 \multicolumn{4}{|c|}{$\tan \beta =2$  $m_t= 170 \:GeV$ }  \\[3pt] 
%\hline  
$A_0, m_0, M_{1/2}$ &300, 300, 80 &  300, 300, 300  &  0,0,300 \\ 
%\hline  
& $\;\;\mu>0\;\;\;\;\mu<0$ & $\;\;\mu>0\;\;\;\;\mu<0$ 
& $\;\;\mu>0\;\;\;\;\mu<0$ \\ 
\hline
\hline
&\multicolumn{3}{|c|}{ $\sqrt{s}=$190GeV} \\  \cline{2-4} 
\multicolumn{1}{|c|}{$\Delta k_{\gamma}$} 
& \verb+ -1.989 -1.783+  & \verb+ -1.793 -1.818+  & \verb+ -1.812 -1.841+ \\ 
\multicolumn{1}{|c|}{$\Delta Q_{\gamma}$}  
& \verb+  0.903  0.297+  & \verb+  0.523  0.525+  & \verb+  0.534  0.537+ \\ 
\multicolumn{1}{|c|}{$\Delta k_{Z}$}      
& \verb+ -2.359 -2.065+  & \verb+  2.209 -2.196+  & \verb+ -2.208 -2.189+ \\ 
\multicolumn{1}{|c|}{$\Delta Q_{Z}$}      
& \verb+ -0.366 -1.197+  & \verb+  0.342  0.360+  & \verb+  0.331  0.354+ \\ 
\hline 
 SM    & 
\multicolumn{3}{|c|}{
$\Delta k_{\gamma} = -2.005, -1.735, -2.118 $ \hfill 
%& & 
$\Delta Q_{\gamma} =0.524,0.530,0.503 $ } \\[3pt] 
 predictions  &
\multicolumn{3}{|c|}{
$\Delta k_{Z} =-1.350,-2.437,-1.404 $ \hfill 
%& &     
$\Delta Q_{Z} =0.507,0.533,0.481$ } \\     
\hline         
\hline         
&\multicolumn{3}{|c|}{ $\sqrt{s}=$500GeV} \\ \cline{2-4} 
\multicolumn{1}{|c|}{$\Delta k_{\gamma}$} 
& \verb+ -0.262 -0.310+  & \verb+ -0.151 -0.207+  & \verb+ -0.191 -0.259+ \\ 
\multicolumn{1}{|c|}{$\Delta Q_{\gamma}$}   
& \verb+  0.150  0.146+  & \verb+  0.030  0.041+  & \verb+  0.056  0.069+ \\ 
\multicolumn{1}{|c|}{$\Delta k_{Z}$}       
& \verb+  0.115  0.238+  & \verb+  0.198  0.191+  & \verb+  0.203  0.198+ \\ 
\multicolumn{1}{|c|}{$\Delta Q_{Z}$}       
& \verb+  0.362  0.256+  & \verb+ -0.407 -0.325+  & \verb+ -0.427 -0.352+ \\ 
\hline 
 SM    & 
\multicolumn{3}{|c|}{
$\Delta k_{\gamma} = -0.250,-0.168,0.046 $ \hfill 
%& & 
$\Delta Q_{\gamma} =0.054,0.057,0.064  $ } \\[3pt] 
 predictions  &
\multicolumn{3}{|c|}{
$\Delta k_{Z} = 0.147, 0.208, -0.036 $\hfill 
%& &     
$\Delta Q_{Z} = 0.057, 0.058, 0.077 $ } \\     
\hline         
\end{tabular}                        
\caption{
MSSM predictions for $\Delta k_{\gamma, Z}$,  
$\Delta Q_{\gamma, Z}$, in units of $g^2/{16\pi^{2}}$, 
for three different inputs of $A_0, m_0, M_{1/2}$ (in GeV), 
at LEP2 and NLC energies.
Both $\mu >0$ and $\mu <0$ cases are displayed. 
The SM  predictions for Higgs masses 
$50, 100$  and $300$ $GeV$ respectively are also displayed. 
}  
\label{tab1}       
%\vspace*{1.5cm}     
\end{center}        
\end{table} 
 
%%%%%%%%%%%%%%%%%%%%%%%%%%%%%%%%%%%%%%%%%%%%%%%%%%%%%%%%%%%%%%%%%%%%%%%%%%%%%%

\vspace*{1cm}  
\noindent 
{\large {\bf Figure Captions}} 
 
\vspace{.4cm} 
\noindent   
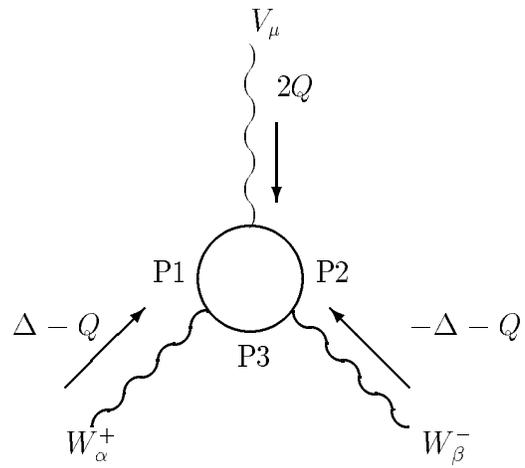
{\bf Figure 1}:\quad The kinematics of the $WWV$ vertex. $P_{1,2,3}$ 
denote internal particle lines.       
 
\vspace{.4cm} 
\noindent  
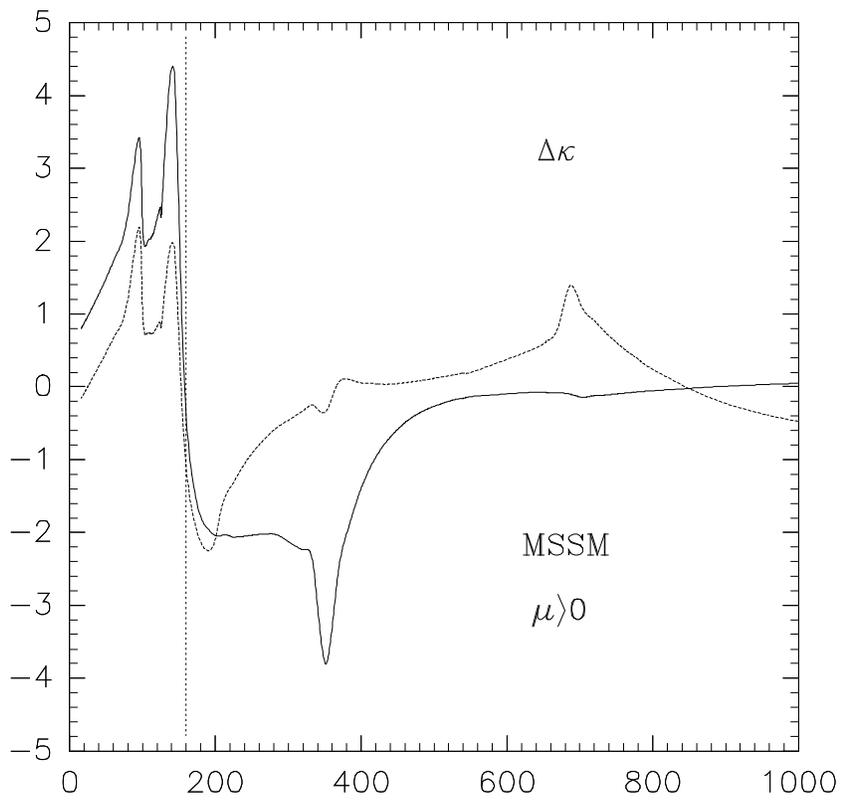
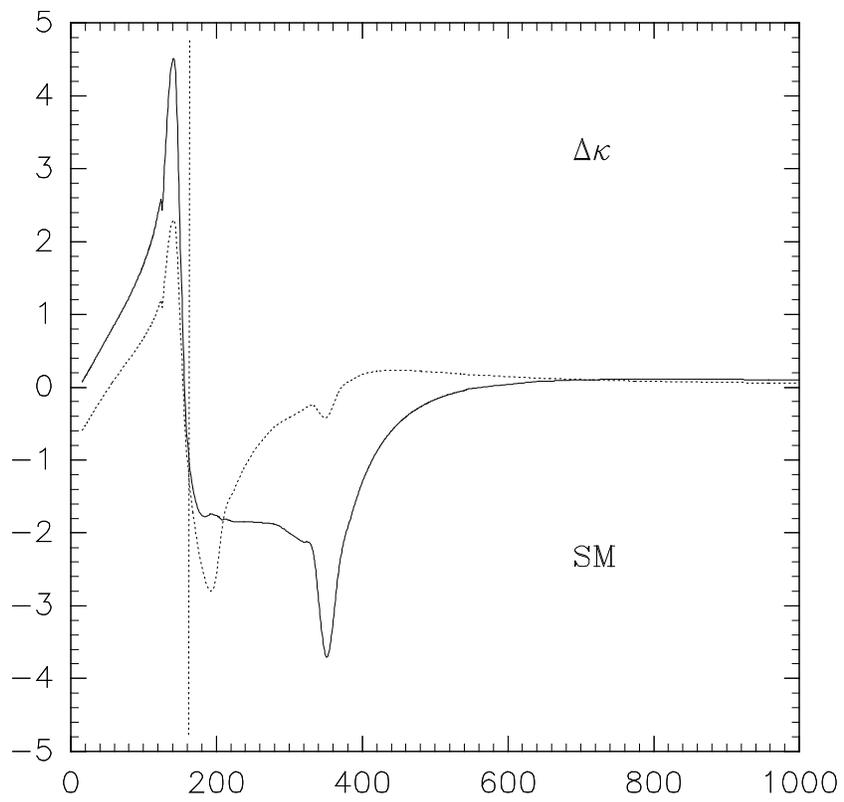
{\bf Figure 2}:\quad {\bf a)} MSSM  predictions for the real part of 
${\Delta k}_{\gamma}$ (solid lines) and  
${\Delta k}_{Z}$, (dashed lines), in units of   
$g^2/{16\pi^{2}}$, as functions of the energy $\sqrt s$. The inputs are  
$(A_0, m_0, M_{1/2})$ $\,=\,$ $(300, 300, 80)$ $GeV$, $\tan \beta \, = \, 2$,   
$m_t \, =\,170 \, GeV$. Only the MSSM case $\mu > 0$ is displayed. 
The vertical dotted line indicates the position of the $WW$ production 
threshold. {\bf b)} SM predictions for $m_t \, =\,170$GeV 
and SM Higgs mass, $m_H \, =\, 100$GeV. 
 
\vspace{.4cm} 
\noindent  
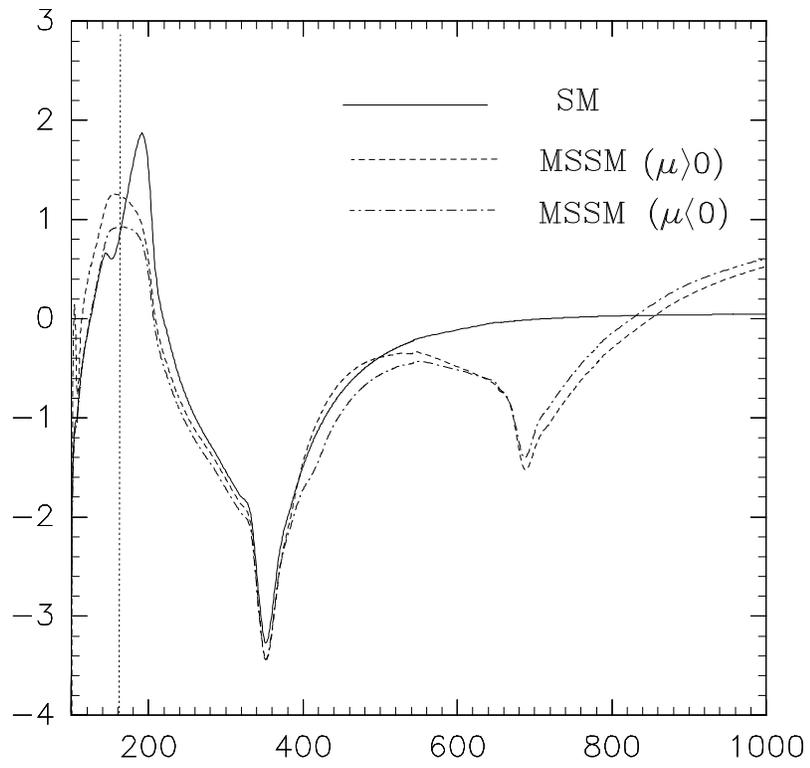
{\bf Figure 3}:\quad $\Delta k_\gamma - 4 Q^2/(4 Q^2-m_Z^2)\Delta k_Z$   
as a function of the energy, 
for the SM (solid line), MSSM with $\mu>0$ (dashed line) and  
MSSM with $\mu<0$ (dashed doted line). The MSSM parameters are as in 
Figure 2. The standard model Higgs mass is taken $100 GeV$.

%%%%%%%%%%%%%%%%%%%%%%%%%%%%%%%%%%%%%%%%%%%%%%%%%%%%%%%%%%%%%%%%%%%%%%%%%%%%%%

\end{document}